\documentclass[fleqn,usenatbib,letters]{mnras}

\usepackage{newtxtext,newtxmath}
\usepackage{hyperref}

\usepackage[T1]{fontenc}

\DeclareRobustCommand{\VAN}[3]{#2}
\let\VANthebibliography\thebibliography
\def\thebibliography{\DeclareRobustCommand{\VAN}[3]{##3}\VANthebibliography}

\usepackage{graphicx}	
\usepackage{amsmath}	

\title[AT2023fhn (the Finch)]{AT2023fhn (the Finch): a Luminous Fast Blue Optical Transient at a large offset from its host galaxy}

\author[A. A. Chrimes et al.]{A. A. Chrimes,$^{1,2}$\thanks{E-mail: ashley.chrimes@esa.int}
P. G. Jonker$^{2,3}$,
A. J. Levan$^{2,4}$,
D. L. Coppejans$^{4}$,
N. Gaspari$^{2}$,
B. P. Gompertz$^{5}$,\newauthor
P. J. Groot$^{2,6,7}$,
D. B. Malesani$^{2,8,9}$,
A. Mummery$^{10}$, 
E. R. Stanway$^{4}$
and K. Wiersema$^{11}$
\\
$^{1}$European Space Agency (ESA), European Space Research and Technology Centre (ESTEC), Keplerlaan 1, 2201 AZ Noordwijk, the Netherlands \\
$^{2}$Department of Astrophysics/IMAPP, Radboud University, PO Box 9010, 6500 GL Nijmegen, The Netherlands \\
$^{3}$SRON, Netherlands Institute for Space Research, Niels Bohrweg 4, 2333 CA, Leiden, The Netherlands \\
$^{4}$Department of Physics, University of Warwick, Gibbet Hill Road, Coventry, United Kingdom \\
$^{5}$Institute of Gravitational Wave Astornomy and School of Physics and Astronomy, University of Birmingham, Birmingham, B15 2TT, United Kingdom \\
$^{6}$Inter-University Institute for Data Intensive Astronomy, Department of Astronomy, University of Cape Town, Private Bag X3, Rondebosch 7701, South Africa \\
$^{7}$South African Astronomical Observatory, P.O. Box 9, 7935 Observatory, South Africa \\
$^{8}$Cosmic Dawn Center (DAWN), Denmark \\
$^{9}$Niels Bohr Institute, University of Copenhagen, Jagtvej 128, 2200 Copenhagen N, Denmark\\
$^{10}$Oxford Astrophysics, Denys Wilkinson Building, Keble Road, Oxford, OX1 3RH, United Kingdom\\
$^{11}$Centre for Astrophysics Research, University of Hertfordshire, Hatfield, AL10 9AB, United Kingdom\\
}

\date{Accepted XXX. Received YYY; in original form ZZZ}

\pubyear{2023}

\begin{document}
\label{firstpage}
\pagerange{\pageref{firstpage}--\pageref{lastpage}}
\maketitle

\begin{abstract}
Luminous Fast Blue Optical Transients (LFBOTs) - the prototypical example being AT\,2018cow - are a rare class of events whose origins are poorly understood. They are characterised by rapid evolution, featureless blue spectra at early times, and luminous X-ray and radio emission. LFBOTs thus far have been found exclusively at small projected offsets from star-forming host galaxies. We present Hubble Space Telescope, Gemini, Chandra and Very Large Array observations of a new LFBOT, AT\,2023fhn. The Hubble Space Telescope data reveal a large offset ($>3.5$ half-light radii) from the two closest galaxies, both at redshift $z\sim0.24$. The location of AT\,2023fhn is in stark contrast with previous events, and demonstrates that LFBOTs can occur in a range of galactic environments.
\end{abstract}




\section{Introduction}
The development of wide-field, high cadence and deep optical surveys in recent years - including the Zwicky Transient Facility \citep[ZTF,][]{2019PASP..131a8002B}, Asteroid Terrestrial-impact Last Alert System \citep[ATLAS,][]{2018PASP..130f4505T}, Panoramic Survey Telescope and Rapid Response System \citep[PanSTARRS,][]{2016arXiv161205560C}, Gravitational-wave Optical Transient Observer \citep[GOTO,][]{2022MNRAS.511.2405S} and Black hole Gravitational-wave Electromagnetic counterpart array \citep[BlackGEM,][]{2016SPIE.9906E..64B}, to name a few - is leading to ever more transient detections in the extremes of parameter space. This trend is set to continue with the Vera Rubin Observatory \citep{2009arXiv0912.0201L}. Such surveys led to the discovery of fast blue optical transients (FBOTs), first identified as a class by \citet{2014ApJ...794...23D} in ZTF. FBOTs rise and fade on timescales of days, and have (early-time) $g$-$r$ colours of -0.3 or bluer. These events also have featureless, black-body-like spectra at early times with inferred temperatures $>10^{4}$\,K \citep{2018MNRAS.481..894P}. It has since become clear that the majority are infant supernovae with low ejecta masses \citep{2018MNRAS.481..894P}, but a small number fade too rapidly to be powered by Ni-56 decay (faster than 0.2-0.3 magnitudes per day), have peak absolute magnitudes rivalling superluminous supernovae ($<-20$), and have accompanying luminous X-ray and radio emission. These bright, multi-wavelength FBOTs have been dubbed luminous-FBOTs \citep[LFBOTS,][]{2022ApJ...932...84M}, the first example of which is AT\,2018cow \citep["the Cow",][]{2018ApJ...865L...3P,2019ApJ...872...18M,2019MNRAS.484.1031P}. Since AT\,2018cow, several other LFBOTs have been discovered (both in real time and archival searches), with varying degrees of multi-wavelength coverage. These include ZTF18abvkwla \citep["the Koala",][]{2020ApJ...895...49H}, CSS161010 \citep{2020ApJ...895L..23C}, ZTF20acigmel \citep["the Camel",][]{2021MNRAS.508.5138P,2022ApJ...926..112B,2022ApJ...932..116H}, AT2020mrf \citep[][]{2022ApJ...934..104Y} and AT\,2022tsd \citep["the Tasmanian Devil",][]{2022TNSAN.199....1H,2023arXiv230601114M}. There are also a number of other lower-confidence candidates \citep[e.g.][]{2022TNSAN.275....1H,2022ApJ...933L..36J,2023TNSAN..41....1P}. Despite the growing number of LFBOT discoveries, these events are intrinsically rare - the volumetric rate of AT\,2018cow-like LFBOTs is estimated to be no more than 0.1 per cent of the local supernova rate \citep{2023ApJ...949..120H}.

The nature of LFBOTs remains unclear. The timescale of their light-curve evolution, X-ray and radio luminosity, late-time UV emission in the case of AT\,2018cow \citep[][]{2022MNRAS.512L..66S,2023MNRAS.519.3785S,2023arXiv230303500C,2023arXiv230807381I}, and preference for star-forming dwarf and spiral hosts have proved challenging to explain with a single self-consistent model. Circumstellar medium interactions around young supernovae are a plausible origin for the early-time spectra and X-ray/radio emission of some FBOTs \citep{2018MNRAS.481..894P,2023ApJ...949..120H}, as well as for the optical polarisation behaviour \citep{2023MNRAS.521.3323M}. However, the peak absolute magnitude, rapid subsequent fading, high radio/X-ray luminosity and peculiar optical and radio polarisation of LFBOTs \citep{2019ApJ...878L..25H,2023MNRAS.521.3323M} require an alternative explanation. Following AT\,2018cow, a few main classes of model emerged. These include central engines born in low-ejecta core-collapse events, powered by black hole accretion or magnetar spin-down \citep[e.g.][]{2019MNRAS.484.1031P,2019ApJ...872...18M}; mergers of stellar-mass black holes and hydrogen-poor stars \citep[e.g.][]{2022ApJ...932...84M}; or the tidal disruption of a main sequence star \citep{2019MNRAS.484.1031P} or white dwarf by an intermediate mass black hole \citep[IMBH,][]{2019MNRAS.487.2505K}. The former is motivated by the rapid light-curve decay and multi-wavelength evolution which severely limits the possible ejecta mass; the latter two also by the timescale - which is too fast for a supermassive black hole (SMBH) tidal disruption event (TDE) - and the weak (initially absent) hydrogen lines in the spectra. Many of these scenarios face challenges. For example, a magnetar central engine can power the early or late-time UV emission in AT\,2018cow, but not both \citep{2023arXiv230303501C}, while the environments of LFBOTS thus far - at small offsets within star-forming dwarfs and spirals, and with high circumstellar densities \citep{2019ApJ...872...18M} - favour a short-lived, massive star progenitor over an IMBH TDE. Further insight will come from similarly detailed studies of other LFBOTs, to establish which features are common to all objects in this class, and to understand the variety among them.

\begin{figure*}
    \includegraphics[width=\textwidth]{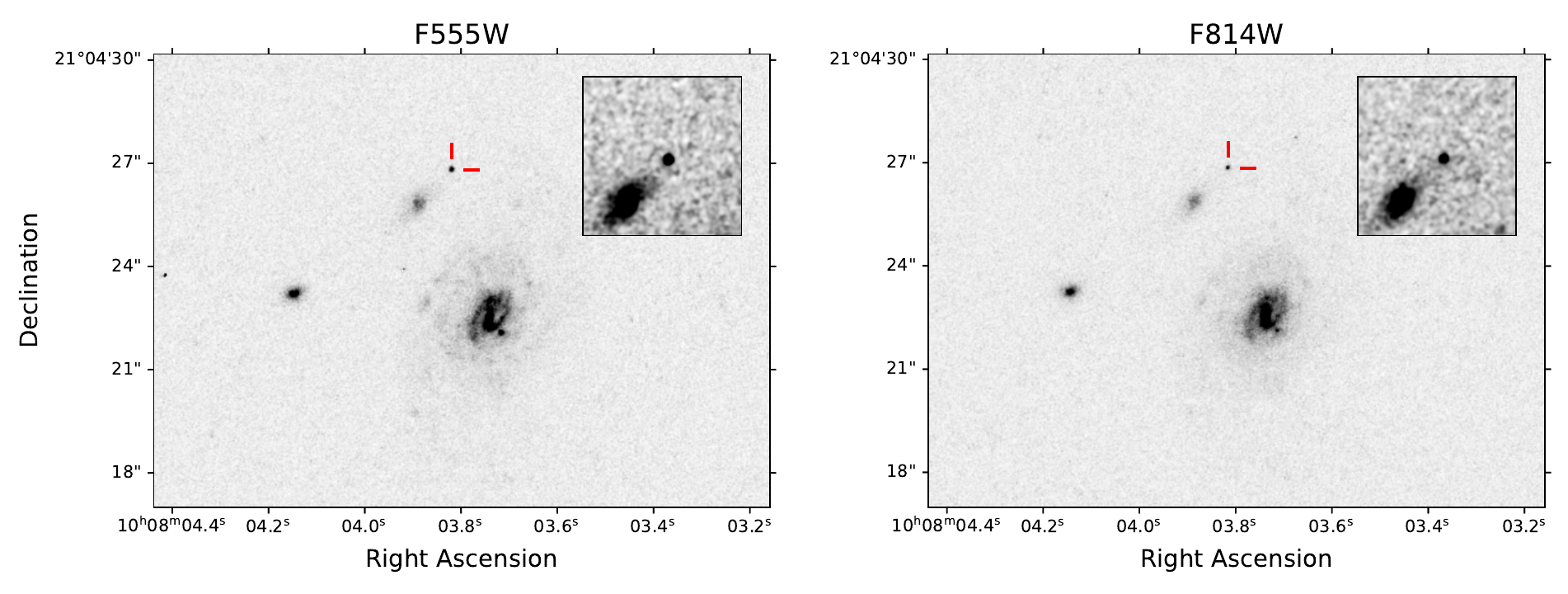}
    \vspace{-0.7cm}
    \caption{{\em HST} images of AT\,2023fhn, indicated by red pointers, and the nearby host galaxy candidates. North is up and east is left in all images. The transient lies at a large offset from both the barred spiral to the south and the dwarf galaxy to the southeast. Smoothed and scaled 3.75$\times$3.75\,arcsec cutouts around AT\,2023fhn are shown in the inset panels. The diffuse emission northwest of the dwarf (satellite) galaxy is an alternative parent stellar population.}
    \label{fig:hst}
\end{figure*}

In this letter, we present multi-wavelength observations of a new LFBOT, AT2023fhn ("the Finch"). The transient is significantly offset from the nearest galaxies, representing a deviation in terms of its environment from previous LFBOTs. This letter is structured as follows. In Section~\ref{sec:discovery} we review how AT\,2023fhn was discovered, and present early-time X-ray and radio observations. Section~\ref{sec:data} presents follow-up observations, including {\it Hubble Space Telescope} ({\em HST}) imaging and Gemini spectroscopy. In Section~\ref{sec:interp} we discuss possible interpretations, and conclusions are drawn in Section \ref{sec:conc}. We adopt a cosmology with H$_{0}=69.6$\,km\,s$^{-1}$\,Mpc$^{-1}$, $\Omega_{\rm m}=0.29$ and $\Omega_{\Lambda}=0.71$ \citep{2006PASP..118.1711W,2014ApJ...794..135B}. Uncertainties are given as 1$\sigma$ unless otherwise stated, and magnitudes are quoted in the AB system \citep{1982PASP...94..586O}.

\section{Discovery and classification}\label{sec:discovery}

\subsection{Early photometry and spectra}
AT\,2023fhn was discovered on 10 Apr 2023 with $m(r)=19.74$ by ZTF \citep{2023TNSTR.775....1F}. The blue colour of $g-r \sim -0.47$ and rapid $\sim$0.2\,mag\,day$^{-1}$ evolution immediately classified AT\,2023fhn as an LFBOT candidate. \citet{2023TNSAN..93....1H} subsequently obtained Gemini GMOS-S spectroscopy of AT\,2023fhn on 19-04-2023 (programme GS-2023A-Q-127), finding a featureless blue spectrum. On 20 Apr 2023 they obtained a spectrum of the nearby spiral galaxy ($\sim$5\,arcsec offset), yielding a redshift of $z\sim0.24$. At this redshift, the earliest ZTF $g$-band (12 Apr 2023) absolute magnitude is -21.5. 

\begin{table}
	\centering
	\caption{VLA flux density upper-limits. These are given as 3 times the local RMS. The third column lists the bandwidth. The final column lists limits on the luminosity, assuming a redshift of $z=0.238$ (see Section \ref{sec:spec}).}
	\label{tab:radio_table}
	\begin{tabular}{cccccc} 
		\hline
		\hline
		  Start date & Freq. & BW & T$_{\rm exp}$ & Upper-limit & Upper-limit \\
            JD-2460056 & GHz & GHz & Min. & $\mu\rm{Jy/beam}$ & 10$^{28}$\,erg\,s$^{-1}$\,Hz$^{-1}$ \\
		\hline
            0.80733 & 1.50 & 1.024 & 35.9 & 130 & 22.5 \\
            0.78309 & 3.00 & 2.048 & 30.0 & 35 & 6.0 \\
            0.76507 & 6.05 & 2.048 & 21.0 & 18 & 3.1 \\
            0.74688 & 10.00 & 4.096 & 21.1 & 18 & 3.1 \\
            0.72090 & 15.02 & 6.144 & 30.1 & 11 & 1.9 \\
            0.69229 & 21.94 & 8.192 & 28.2 & 17 & 2.9 \\
            0.66552 & 32.94 & 8.192 & 25.4 & 25 & 4.3 \\
		\hline
	\end{tabular}
\end{table}

\subsection{X-ray and radio observations}
We triggered {\it Chandra X-ray Observatory} observations (PI: Chrimes; program 24500143; Obs ID 26624), which were obtained on 25 Apr 2023 (06:58:08 -- 15:46:51\,UT). The faint-mode ACIS-S exposure lasted 30\,ks. The data were reduced and analysed with standard {\sc ciao} (v4.13, caldb v4.9.3) procedures including reprocessing, filtering and source measurement with {\sc srcflux}. Assuming a power-law with a photon index $\Gamma=2$ \citep{2018MNRAS.480L.146R,2023arXiv230601114M}, the unabsorbed source flux after correction for the Galactic neutral hydrogen column density of $NH=2.4\times10^{20}$cm$^{-2}$ \citep{2005A&A...440..775K} is 7.6$^{-1.8}_{+2.2} \times 10^{-15}$\,erg\,cm$^{-2}$\,s$^{-1}$ (0.5-7.0\,keV). At the redshift of the spiral, this corresponds to a luminosity of 1.3$^{-0.3}_{+0.4} \times$10$^{42}$\,erg\,s$^{-1}$, comparable to other LFBOTs at the same epoch \citep{2018MNRAS.480L.146R,2019ApJ...872...18M,2019MNRAS.487.2505K,2020ApJ...895L..23C,2022ApJ...926..112B,2022ApJ...934..104Y,2023arXiv230601114M}.

Early radio observations (within a few weeks of discovery) produced non-detections, including a 10\,GHz Northern Extended Millimeter Array upper limit of $2\times 10^{29}$\,erg\,s$^{-1}$\,Hz$^{-1}$ on the luminosity \citep{2023TNSAN.174....1H}, and upper limits from our own programme (SC240143, PI: Chrimes) on the Karl G. Jansky Very Large Array (VLA). We observed AT\,2023fhn on 22 Apr 2023 ($\approx12$ days post detection) in standard phase-referencing mode using 3C286 as a flux density and bandpass calibrator, with J1014+2301 and J1016+2037 as complex gain calibrators. The observations were calibrated using the VLA Calibration Pipeline 2022.2.0.64 in CASA version 6.4.1 with additional manual flagging. We imaged the data using the task {\sc tclean} in CASA with Briggs weighting with a robust parameter of 1. No significant emission was detected at the source location. We provide the upper-limits in Table \ref{tab:radio_table}. These early-time non-detections are consistent with the behaviour of previous LFBOTs. The transient was subsequently detected with the VLA on 15 Jun 2023 \citep{2023TNSAN.174....1H} with luminosity $7.6\times 10^{28}$\,erg\,s$^{-1}$\,Hz$^{-1}$ (at 10\,GHz), also similar to other LFBOTs at the same epoch \citep[e.g.][]{2019ApJ...872...18M,2020ApJ...895L..23C}. The rapid evolution (timescale of a few days) and peak optical absolute magnitude of -21.5 places AT\,2023fhn firmly within the LFBOT region of timescale/peak luminosity parameter space \citep[see Figures 3 and 14 of][]{2023ApJ...949..120H}. Along with the hot featureless optical spectrum, X-ray and radio detections, AT\,2023fhn is unambiguously identified as a new AT\,2018cow-like LFBOT.

\section{Follow-up observations}\label{sec:data}

\subsection{Hubble Space Telescope Imaging}

\subsubsection{Data reduction and photometry}
{\em HST} WFC3/UVIS observations were taken with the F555W and F814W filters on 17 May 2023 (PI: Chrimes; proposal ID 17238). Three 364\,s exposures with sub-pixel dithers were taken in each filter. The F555W exposures began 09:02:23 and ended 09:23:41\,UT, the F814W exposures began 09:25:31 and ended 09:48:13\,UT. The {\sc $\_$flc} images were combined using {\sc astrodrizzle}\footnote{Part of {\sc drizzlepac}, \url{http://drizzlepac.stsci.edu/}} \citep{2002PASP..114..144F}, with {\sc pix$\_$frac} = 0.8 and a final pixel scale of 0.025\,arcsec\,pixel$^{-1}$. The transient is clearly identified in the reduced images, as shown in Figure~\ref{fig:hst}. Two adjacent galaxies are fully resolved: a barred spiral to the south and a dwarf/irregular to the southeast. These galaxies have Sloan Digital Sky Survey (SDSS) data release 16 \citep{2020ApJS..249....3A} IDs SDSS\,J100803.73$+$210422.5 and SDSS\,J100803.87+210425.8. We perform photometry on AT\,2023fhn with three methods. The first two use standard {\sc photutils} aperture photometry procedures in python \citep{larry_bradley_2021_4624996}, but the background level is calculated in two ways. The first uses the {\sc MedianBackground} estimator (using the whole image for the estimate). The second uses an annulus around the source (inner and outer radii of 1.5 and 4 times the aperture radius, and pixel values in the annulus clipped at $\pm 3 \sigma$). For each of these background estimations, two aperture sizes are used - 0.2 and 0.4\,arcsec - with the appropriate aperture corrections for F555W and F814W applied\footnote{\url{https://hubblesite.org/sites/www/home/hst/instrumentation/wfc3/data-analysis/photometric-calibration/uvis-encircled-energy}}. AB magnitudes are derived from the {\sc photflam} and {\sc photplam} header values and the published conversion procedures\footnote{\url{https://hst-docs.stsci.edu/wfc3dhb/chapter-9-wfc3-data-analysis/9-1-photometry}}. For the third method we use {\sc dolphot} \citep[v2.0,][]{2000PASP..112.1383D}. {\sc dolphot} performs PSF photometry on each {\sc $\_$flc} image separately; these measurements are combined to give the reported value and its error. {\sc dolphot} provides instrumental magnitudes in the Vega system, but we instead report AB magnitudes using conversions calculated with {\sc stsynphot} \citep{2020ascl.soft10003S}. Magnitude measurements for each combination of filter and methodology are given in Table~\ref{tab:phot_table}. Smaller apertures and annulus background subtraction results in fainter magnitudes, indicative of the presence of diffuse emission around the transient (as can be seen in Figure \ref{fig:hst}, see insets).

\begin{table}
	\centering
	\caption{{\em HST} magnitudes $m$, and their uncertainties $\delta m$, for AT\,2023fhn. In both filters, three photometry methods are listed - aperture photometry with median background estimation, aperture photometry with annulus background estimation, and {\sc dolphot}. For the non-{\sc dolphot} measurements, two aperture sizes (and hence enclosed energy corrections) are listed.}
	\label{tab:phot_table}
	\begin{tabular}{lccccr} 
		\hline
		\hline
		Filter & Method & Background & Aperture & m & $\delta$m \\
		\hline
            F555W & {\sc photutils} & Median & 0.2$''$ & 24.31 & 0.02 \\
            F555W & {\sc photutils} & Annulus & 0.2$''$ & 24.38 & 0.02 \\
            F555W & {\sc photutils} & Median & 0.4$''$ & 24.13 & 0.03 \\
            F555W & {\sc photutils} & Annulus & 0.4$''$ & 24.30 & 0.02 \\
            F555W & {\sc dolphot} & -- & PSF & 24.57 & 0.01 \\
            
            F814W & {\sc photutils} & Median & 0.2$''$ & 24.17 & 0.03 \\
            F814W & {\sc photutils} & Annulus & 0.2$''$ & 24.27 & 0.02 \\
            F814W & {\sc photutils} & Median & 0.4$''$ & 23.94 & 0.04 \\
            F814W & {\sc photutils} & Annulus & 0.4$''$ & 24.11 & 0.03 \\
            F814W & {\sc dolphot} & -- & PSF & 24.45 & 0.07 \\
		\hline
	\end{tabular}
\end{table}

\subsubsection{Galaxy offsets and enclosed flux radii}
The sky-projected spatial offset of a transient from its host is a key piece of information for understanding its origin. Host-normalised offsets, offsets divided by the half-light radius of the host, are widely used in the literature (see Figure \ref{fig:offsets}) as they account for the projected extent of the host galaxy. In order to measure the offsets and host-normalised offsets of AT\,2023fhn from the two nearby galaxies, we measure their centroids and half-light radii $r_{50}$ (from Petrosian profile fitting) using the python package {\sc statmorph} \citep{2019MNRAS.483.4140R}. We require objects to have at least 5 adjacent pixels, each $>$1\,$\sigma$ above the background. The resultant segmentation maps are convolved with a uniform filter of size 10 pixels and these filtered segmentation maps are used to identify objects by requiring values $>0.5$. Enclosed flux measurements are not restricted to the galaxy-associated pixels identified with this method; flux is measured out to r$_{\rm max}$ which extends beyond the segmentation area to the faint outer regions \citep[further than twice then Petrosian radius, for details see][]{2019MNRAS.483.4140R}. We note that the transient lies outside the pixels selected as associated with the galaxy in both cases. Segmentation maps, radial light profiles in the direction of the transient, and {\sc statmorph} S{\'e}rsic fits for the two galaxies in each filter, are provided in the associated github repository\footnote{\url{https://github.com/achrimes2/Finch}}. 

At $z=0.238$ - the redshift of the spiral (and its satellite, see Section~\ref{sec:spec}) - the physical scale is 3.80\,kpc\,arcsec$^{-1}$. From the centre of the spiral, the projected offset of AT\,2023fhn $\delta r$ is $16.51\pm0.09$\,kpc. From the centre of the satellite, the offset is $5.35\pm0.06$\,kpc (uncertainties as described below). The non-parametric half-light radius r$_{50}$ (enclosing 50 per cent of the flux, $r_{50}$) is measured to be $4.5\pm0.2$\,kpc in F555W for the spiral. Given the satellite's ellipticity of 0.4 and the orientation of AT\,2023fhn, we take r$_{50}$ along the semi-major axis, which is $1.48\pm0.10$\,kpc in F555W. In F814W, these values are $3.90\pm0.13$\,kpc and $1.29\pm0.10$\,kpc, respectively. This corresponds to host-normalised offsets ($r_{\rm n}= \delta r$/$r_{50}$) of $3.7\pm0.2$ and $3.6\pm0.2$ in F555W, while in F814W,  $r_{\rm n} = 4.25\pm0.14$ and $4.1\pm0.3$ (for the spiral and satellite respectively). The quoted offset uncertainties are the quadrature sum of the transient positional uncertainty (given by FWHM/(2.35$\times$SNR), where FWHM is the full-width at half-maximum and SNR the signal-to-noise ratio) and the uncertainty on the galaxy centroids ($x_{\rm c}$,$y_{\rm c}$). The centroid uncertainties are calculated by re-sampling the input {\sc $\_$flc} image set 100 times using their [ERR] extensions, re-drizzling each re-sampled set, and measuring the morphological properties with {\sc statmorph} on each iteration of the re-drizzled image \citep[see][]{2017MNRAS.467.1795L,2019MNRAS.486.3105C}. The mean and standard deviation of the resultant $x_{\rm c}$, $y_{\rm c}$ and r$_{50}$ distributions are used, along with the AT\,2023fhn positional uncertainties, to calculate the values and their uncertainties quoted above.

\subsubsection{Search for underlying and extended emission}\label{sec:emiss}
\begin{figure}
    \centering
    \includegraphics[width=0.8\columnwidth]{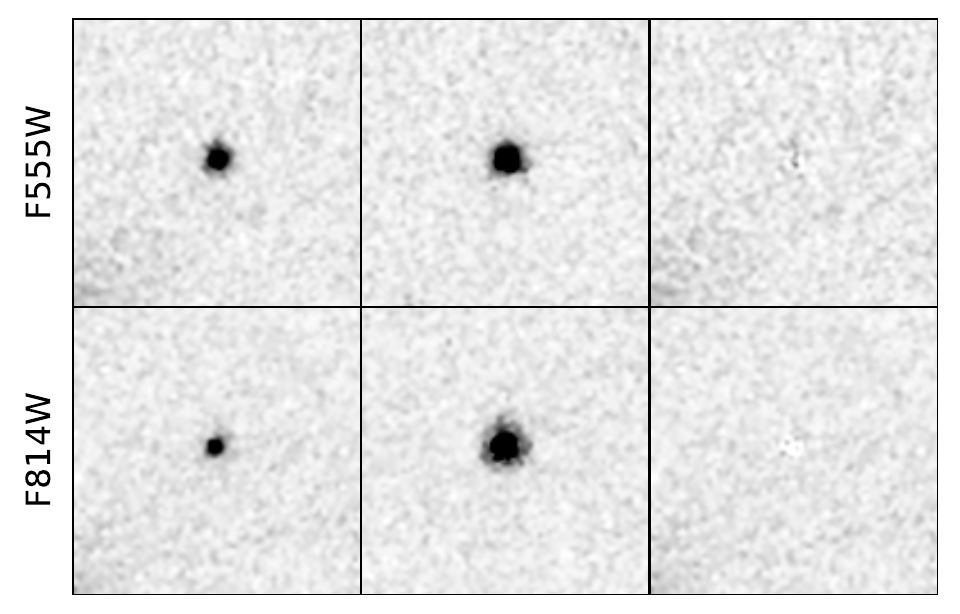}
    \vspace{-0.3cm}
    \caption{Subtraction of a reference star at the location of AT\,2023fhn. The 2$\times$2\,arcsec cutouts show the transient (left), the reference star (middle) and the residual (right), after interpolating onto a finer pixel scale and subtraction of the shifted and vertically scaled reference star. The emission is consistent with being a point source.}
    \label{fig:pointsource}
\end{figure}

Given the apparently isolated location of AT\,2023fhn, it is prudent to search for any underlying (extended) emission at the transient location, such as a knot of star formation, cluster or background galaxy. To establish whether the emission is unresolved, we first select a reference point source in the image (the object at coordinates $\alpha = 10$h08m03.13s, $\delta = +$21d04m22.8s). Cutouts around AT\,2023fhn and the reference star are interpolated onto a pixel grid with twice the resolution (enabling sub-pixel shifts), before subtraction of the reference image from the one containing AT\,2023fhn. The reference is scaled in peak flux and shifted in $x$,$y$ to minimize the standard deviation at the location of AT\,2023fhn in the residual image. The transient, reference and residual images are shown in Figure~\ref{fig:pointsource}. To determine if the residuals are consistent with a clean point source subtraction, we perform {\sc photutils} aperture photometry (with an annulus) as described above. 
No significant residual flux is detected, demonstrating that any underlying (non-transient) source contributing significantly to the flux must be precisely co-located and also unresolved (the physical scale at this distance is 95\,pc\,pixel$^{-1}$). Making use of BPASS \citep[Binary Population and Spectral Synthesis v2.2,][]{2017PASA...34...58E,2018MNRAS.479...75S} synthetic spectra, we calculate the maximum mass of a stellar cluster which can be present at the location of AT\,2023fhn, without exceeding the observed luminosity in either F555W or F814W. We find that the maximum possible mass of an unresolved cluster rises with population age, from $3\times10^{6}$M$_{\odot}$ at 10$^{6}$\,yr to $\sim10^{9}$M$_{\odot}$ at 10$^{10}$\,yr. Therefore, the presence of a typical stellar cluster - at any age - cannot be ruled out. To search for extended emission, we smooth the images with a Gaussian filter ($\sigma=1.5$) and scale them to show diffuse background light. The inset panels of Figure \ref{fig:hst} show cutouts of the smoothed and scaled images. Faint emission can be seen extending northwest of the satellite, plausibly a tidal stream. The surface brightness near the transient location (measured in a 1\,arcsec radius around AT\,2023fhn) is 25.2\,mag\,arcsec$^{-2}$ in F555W and 24.6\,mag\,arcsec$^{-2}$ in F814W.

\begin{figure}
    \includegraphics[width=\columnwidth]{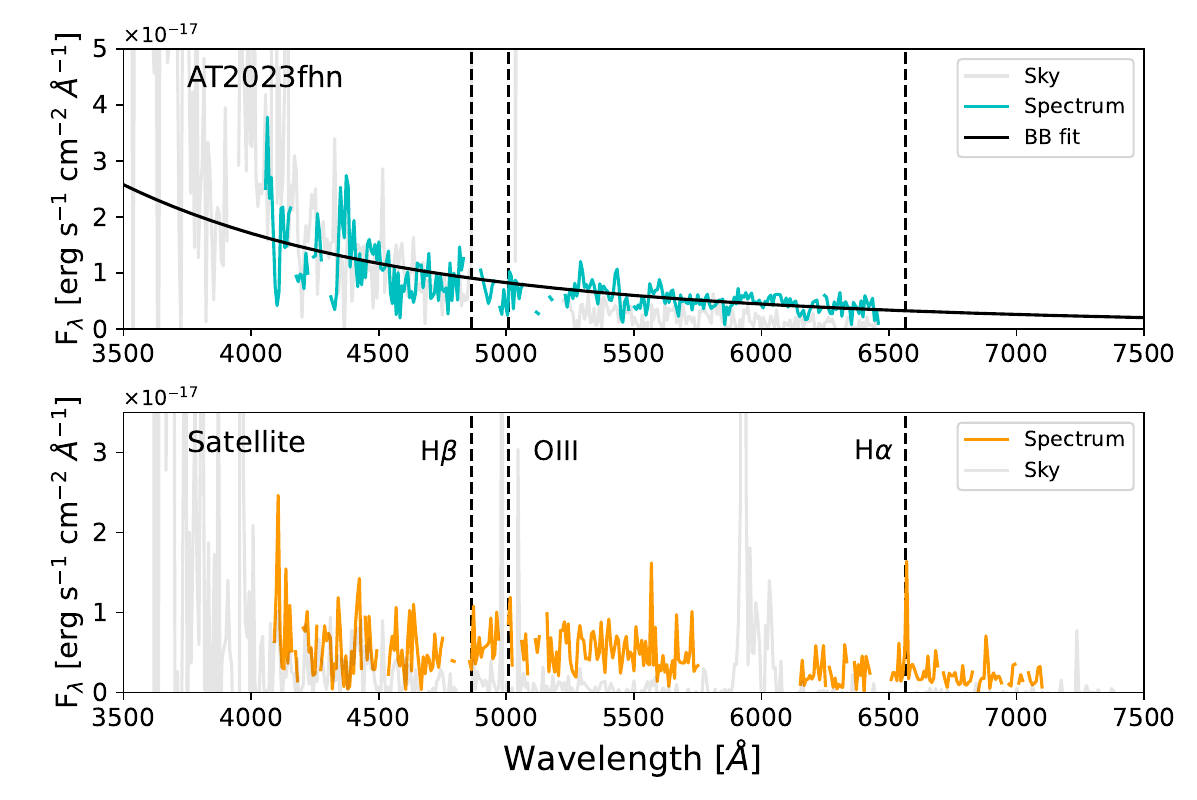}
    \vspace{-0.7cm}
    \caption{Upper panel: the background-subtracted spectrum of AT\,2023fhn obtained with Gemini/GMOS-S on 22/23 Apr 2023, $\sim$10 rest-frame days post-discovery, and shifted into the transient rest-frame. A black-body fit returns $T=24.8^{+2.4}_{-2.3}$\,$\times10^{3}$\,K. Background traces are shown in grey. Lower panel: a spectrum of the satellite galaxy. A robust detection of the H$\alpha$ emission line at $z=0.238\pm0.004$ confirms an association with the adjacent spiral.}
    \label{fig:spectra}
\end{figure}

\begin{figure*}
    \includegraphics[width=\textwidth]{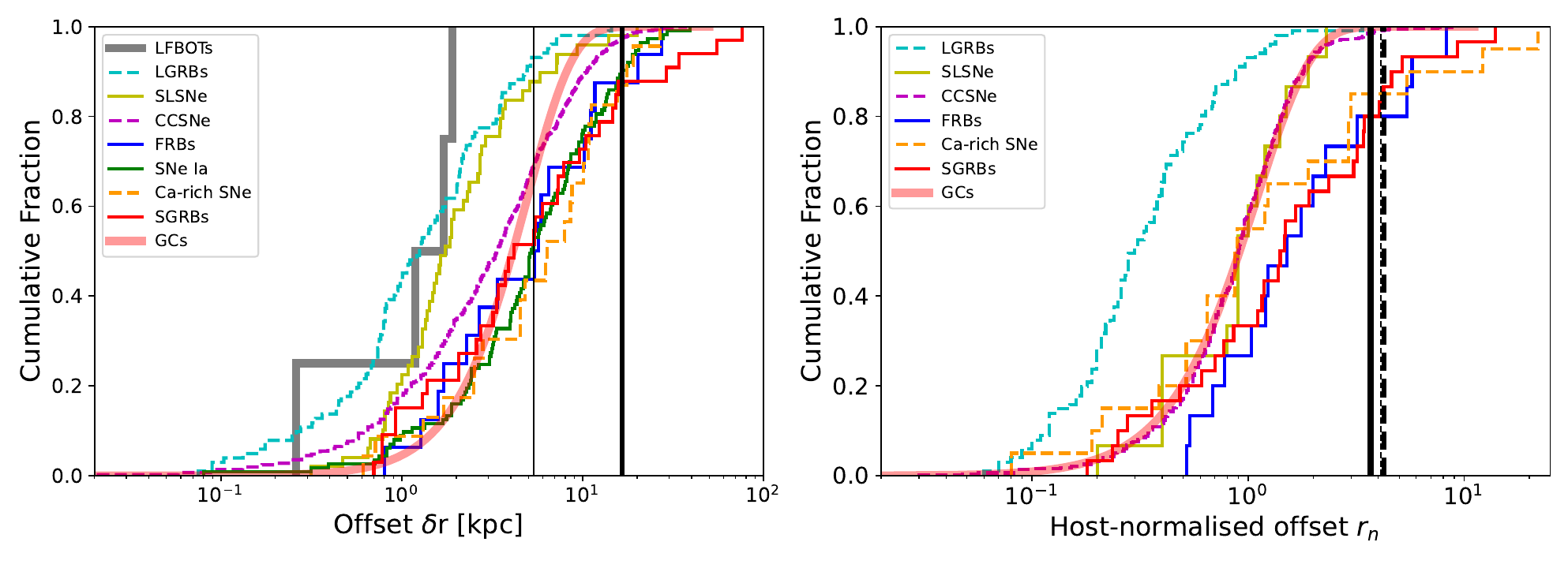}
    \vspace{-0.7cm}
    \caption{The cumulative offset and host-normalised offset distributions of a variety of transients, and the offset of AT\,2023fhn from the spiral (thick black vertical lines) and its satellite (narrow vertical lines) - solid lines represent F555W, dashed lines F814W. The four previous LFBOT offsets are from \citet[][the Cow]{2018ApJ...865L...3P}, \citet[][the Koala]{2020ApJ...895...49H}, \citet[][CSS161010]{2020ApJ...895L..23C} and \citet[][AT\,2020mrf]{2022ApJ...934..104Y}. The comparison distributions are from \citet[][LGRBs]{2016ApJ...817..144B,2017MNRAS.467.1795L}, \citet[][SLSNe]{2015ApJ...804...90L,2021ApJS..255...29S}, \citet[][CCSNe]{2012ApJ...759..107K,2021ApJS..255...29S}, \citet[][FRBs]{2022AJ....163...69B}, \citet[][type Ia SNe]{2013Sci...340..170W}, \citet[][Ca-rich SNe]{2017ApJ...836...60L,2020ApJ...905...58D} and \citet[][SGRBs]{2022ApJ...940...56F}. Also shown is the globular cluster (GC) offset distribution around M81 \citep{2022MNRAS.509..180L}.
    }
    \label{fig:offsets}
\end{figure*}

\subsection{Gemini spectroscopy}\label{sec:spec}
We obtained two epochs of Gemini/GMOS-S spectroscopy on 22/23 Apr 2023 and 12 May 2023, $\sim$10 and $\sim$26 days post discovery respectively (PI: Chrimes, programme GS-2023A-DD-102). The first epoch consisted of 4$\times$500s exposures with the R400 grating, 1\,arcsec slit width and two central wavelengths (two exposures at 520\,nm and two at 565\,nm). The second epoch consisted of 4$\times$1845s exposures with the R400 grating, 1\,arcsec slit and central wavelength 675\,nm. Data reduction was performed using the python package {\sc dragons} \citep{2019ASPC..523..321L}. Associated arcs, flats and bias frames were taken as part of the programme. Sky lines and unusable regions (e.g. due to the amplifier 5 failure\footnote{\url{https://www.gemini.edu/node/21963}}) are manually masked. We bin the pixels by a factor of 6 along the wavelength axis to increase the signal-to-noise ratio, and combine the 520\,nm and 565\,nm centred spectra by taking the mean where they overlap. We correct for Galactic extinction by adopting $E(B-V)=0.025$ \citep{2011ApJ...737..103S}, and calculate the extinction at each wavelength with the python {\sc extinction} \citep{barbary_kyle_2016_804967} module assuming $R_{\rm V}=3.1$. For flux calibration, spectro-photometric standard stars observed with the closest-matching set-up were found in the Gemini archive. For the 525\,nm data we use spectra of EG274 (programme GS-2023A-FT-205), for the 565\,nm data we use LTT6248 (GS-2022A-Q-315) and for the 675\,nm data we use LTT1020 (GS-2022B-Q-126). The final extinction-corrected spectra are plotted in Figure \ref{fig:spectra}. 

In our first epoch of spectroscopy (22/23 Apr), AT\,2023fhn is detected as shown in Figure \ref{fig:spectra}. Fitting a black-body to the Galactic extinction-corrected, rest-frame spectrum yields a temperature of $24.8^{+2.4}_{-2.3}$\,$\times10^{3}$\,K ($\chi^{2}_\nu=3.66$ with 282 degrees of freedom, where uncertainties are derived from the local standard deviation of the spectrum). This compares with a temperature of $17.5^{+1.2}_{-1.0}$\,$\times10^{3}$\,K derived from FORS2 photometry taken on the following night \citep{2023TNSAN.101....1W}. The large $\chi^{2}_\nu$ is likely due to correlated, systematic errors (e.g. from imperfect flux calibration) that have not been accounted for. A power-law produces a fit of similar quality - taking F$_{\lambda} \propto \nu^{2-\beta}$, we find a best-fit power-law index $\beta = -1.24^{+0.06}_{-0.09}$, with $\chi^{2}_\nu=3.63$. Nevertheless, temperatures of $\sim$20\,$\times10^{3}$\,K are comparable to AT\,2018cow, which had a black-body temperature of 19.3$^{+0.7}_{-0.8}$\,$\times10^{3}$\,K at a similar rest-frame epoch \citep{2018ApJ...865L...3P}. No correction for host-intrinsic extinction has been made, however as revealed in the {\em HST} imaging, the transient appears to be far away from any significant sources of dust, as it lies outside the bulk of the optical light of both nearby galaxies. In the second epoch of spectroscopy (12 May) the transient had faded sufficiently to result in a non-detection, with an upper limit on H$\alpha$ emission at its location (taking an aperture with the same radius as the seeing) of $< 1.2\times10^{-16}$\,erg\,s$^{-1}$\,cm$^{-2}$. The slit was also placed on the edge of the satellite galaxy. From the centroid and width of the H${\alpha}$ line, we derive a redshift $z=0.238\pm0.004$, consistent with the spiral redshift of $\sim0.24$ reported by \citet{2023TNSAN..93....1H}, and backing up the satellite interpretation for this galaxy. We have adopted $z=0.238$ for all relevant calculations in this letter.

\section{Discussion}\label{sec:interp}
All published LFBOTS to date have occurred in star-forming dwarfs \citep[the Koala, CSS161010, the Camel, AT\,2020mrf,][]{2020ApJ...895...49H,2020ApJ...895L..23C,2021MNRAS.508.5138P,2022ApJ...934..104Y} or spirals \citep[the Cow,][]{2018ApJ...865L...3P,2020MNRAS.495..992L}. AT\,2023fhn also has a star-forming host, assuming one of the spiral or dwarf  (both are strong H$\alpha$ emitters) is the galaxy of origin. However, in contrast with LFBOTs so far, it lies far away from the bulk of the host light for either choice of host galaxy. Such offsets are atypical for core-collapse transients due to the short (10-100\,Myr lifetimes) of the progenitor stars. Figure \ref{fig:offsets} compares the physical projected offsets and host-normalised offsets of a range of transients compiled from the literature, including long gamma-ray bursts (LGRBs), short gamma-ray bursts (SGRBs), superluminous supernovae (SLSNe), other core-collapse supernovae (CCSNe), fast radio bursts (FRBs), Ca-rich and type Ia SNe. The host offsets of four previous LFBOTs are also shown ($r_{n}$ values were not reported for these events). AT\,2023fhn lies much further out from its host than other LFBOTs to date. To quantify this, we randomly draw 5 (the number of LFBOTs with host offset measurements in Fig. \ref{fig:offsets}) offsets from the \citet{2021ApJS..255...29S} CCSN distribution 10$^{4}$ times, and calculate the frequency with which at least one of these lies at 5.35 (16.51)\,kpc or greater (for the satellite and spiral respectively). For the satellite, this occurs in 85 per cent of random draws, for the spiral it occurs in 13 per cent. In terms of host-normalised offset, only $\sim$1 per cent of CCSNe occur at higher offsets than AT\,2023fhn. In all 4 combinations of filter and galaxy choice, the transient lies outside the pixels selected as associated with the galaxies, therefore (by definition) the transient will have a fraction of light \citep{2006Natur.441..463F} value F$_{\rm light} = 0$ in both filters. This is unlikely but not unprecedented for core-collapse events; a few per cent of CCSN have F$_{\rm light} = 0$ \citep{2010MNRAS.405...57S}. Therefore, a core-collapse origin cannot be ruled out. 

If originating at a lower offset, time-of-travel arguments require a massive star with velocity $\gtrsim$50/350\,km\,s$^{-1}$ for the spiral/satellite, assuming a long-lived 100\,Myr-old progenitor \citep{2019MNRAS.482..870E} and an origin at $\sim$r$_{50}$. Only a small fraction of massive stars have such high velocities \citep[e.g.][]{2000ApJ...544..437P,2005A&A...437..247D,2011MNRAS.414.3501E,2019A&A...624A..66R,2023MNRAS.522.2029C}. The delayed mergers of compact objects can also achieve high offsets (i.e. SGRBs), but the luminosity, spectra and rapid evolution of LFBOTs effectively rule out an association with even the most extreme of these transients \citep[e.g.][]{2011ApJ...734...96K,2022MNRAS.516.4949S}. Since no spectroscopic redshift for the transient has been measured, we consider the probability of a chance alignment P$_{\rm chance}$ between AT\,2023fhn and the two galaxies \citep[following][]{2002AJ....123.1111B,2010ApJ...722.1946B}. P$_{\rm chance}$ is calculated using SDSS DR16 $r$-band magnitudes for the spiral and satellite, which are $18.94\pm0.02$ and $22.61\pm0.14$, respectively. For the spiral we find P$_{\rm chance}=0.78$ per cent, and for the satellite P$_{\rm chance}=1.38$ per cent.  Therefore, AT\,2023fhn is likely associated with one of the two galaxies. As shown in the inset panels of Figure \ref{fig:hst}, the progenitor may have originated in a faint tidal stream or spiral arm. Based on our early-time radio and H$\alpha$ upper limits (Sections \ref{sec:discovery} and \ref{sec:spec}), and using the star formation rate (SFR) calibrations of \citet{2011ApJ...737...67M}, we derive 3\,$\sigma$ upper limits on the underlying SFR at the location of AT\,2023fhn of $\sim$6\,M$_{\odot}$yr$^{-1}$ (at 6.05\,GHz, the strongest radio constraint) and $\sim$0.1\,M$_{\odot}$yr$^{-1}$ (H$\alpha$). The F555W (rest-frame $\sim$B-band) surface brightness of 25.2\,mag\,arcsec$^{-2}$ (Sec. \ref{sec:emiss}) is among the faintest $\sim$2 per cent of ($u$-band) local surface brightnesses for CCSNe \citep[][]{2012ApJ...759..107K}. Unless the population is extremely young, adjusting for the $B$-band/$u$-band discrepancy would give an even fainter surface brightness (due to lower flux blue-wards of the Balmer break). An IMBH TDE explanation requires an underlying cluster, since a dense stellar environment is necessary to make encounters likely \citep[e.g.][]{2023arXiv230307375Y}. As shown in Section \ref{sec:emiss}, a cluster at the location of AT\,2023fhn cannot be ruled out. At $z\sim0.24$, even the brightest and largest globular clusters (GCs) would have optical apparent magnitudes of $\sim$30 - far fainter than the source in the {\em HST} images - and angular extents too small to be resolved \citep{2010arXiv1012.3224H}. Finally, we compare the offset of AT\,2023fhn from the spiral with the distribution of GCs around M81 (which has a similar physical size and morphology), using the S{\'e}rsic distribution of \citet{2022MNRAS.509..180L} \citep[see also][]{1995AJ....109.1055P}. The GC offsets, and distribution normalised by the F555W half-light radius of the spiral, are shown in Figure \ref{fig:offsets}. Only 0.5 per cent of GCs occur at the offset of AT\,2023fhn or higher. While unlikely based on this statistic, the lack of strong photometric constraints mean that an origin in a globular cluster is also not ruled out.

\section{Conclusions}\label{sec:conc}
In this letter, we have presented {\em HST}, Gemini, Chandra and VLA observations of AT\,2023fhn, the first LFBOT to lie at a large offset from its host galaxy. Although the location is more representative of other transient types, given the offset, local surface brightness, limit on star-formation and constraints on an underlying cluster, we cannot rule out a massive star progenitor. Likewise, a tidal disruption event in a unseen cluster cannot be ruled out. Environmental studies are needed for a population of LFBOTs to determine if AT\,2023fhn is a significant outlier. Late-time imaging will put further constraints on the underlying stellar population, while detailed modelling of the spectra and multi-wavelength light-curve is needed to reveal more about the origin of this enigmatic transient. 

\section*{Acknowledgements}
This work is part of the research programme Athena with project number 184.034.002, which is (partly) financed by the Dutch Research Council (NWO). This research has made use of computing facilities provided by the Scientific Computing Research Technology Platform of the University of Warwick. Observations analysed in this work were taken by the NASA/ESA Hubble Space Telescope under program 17238. This research has made use of software provided by the Chandra X-ray Center (CXC) in the application of the CIAO package \citep{2006SPIE.6270E..1VF}. The National Radio Astronomy Observatory is a facility of the National Science Foundation operated under cooperative agreement by Associated Universities, Inc. Based on observations obtained at the international Gemini Observatory, a program of NSF’s NOIRLab, which is managed by the Association of Universities for Research in Astronomy (AURA) under a cooperative agreement with the National Science Foundation on behalf of the Gemini Observatory partnership. Finally, we thank the anonymous referee for their helpful feedback on this manuscript.

\section*{Data Availability}
The data used are available upon request. Scripts and parameter files are available at \url{https://github.com/achrimes2/Finch}.





\bibliographystyle{mnras}
\bibliography{finch_hst_paper} 








\bsp	
\label{lastpage}
\end{document}